\documentstyle[11pt]{article}
\def\v1{\vspace{1cm}}
\def\be{\begin{equation}}
\def\ee{\end{equation}}
\def\bc{\begin{center}}
\def\ec{\end{center}}

\newcommand{\bt}{\begin{tabular}}
\newcommand{\et}{\end{tabular}}

\newcommand{\bea}{\begin{eqnarray}}
\newcommand{\eea}{\end{eqnarray}}

\begin{document}

\title
{\bf Relativity theory of clocks and rulers}

\author{Marek Pawlowski
\\ Soltan Institute for Nuclear Studies, Warsaw, Poland \\ e-mail:
pawlowsk@fuw.edu.pl} \maketitle

\date{\empty}

\begin{abstract}
Special Relativity (SR) kinematics is derived from very intuitive
assumptions. Contrary to standard Einstein's derivation, no light
signal is used in the construction nor it is assumed to exist.
Instead we postulate the existence of two equivalence classes of
physical objects: proportional clocks and proportional rulers.
Simple considerations lead to Lorentz kinematics as one of three
generic cases. The Lorentz case is characterized by the maximal
relative speed of physical objects. The two others are the Galilean
and the Euclidean cases.

\end{abstract}

\section{Preliminaries}

Writing this paper was inspired by the recent interest to so called
Double Special Relativity (DSR) theories \cite{dsr}. Their authors
speculate on the possible modification of standard dispersion
relation and on the fundamental frequency dependence of the
velocity of light. The present paper is not affirmative or critical
to those theories. Our aim is to show that SR can be derived from
the set of intuitive assumptions which does not ascribe a special
role to light signals. This result seams to be interesting by
itself. Only as a byproduct we can point out that consequently the
DSR physics of \textsl{c} can be studied without basic SR
restrictions.

\bigskip

Let us specify several elementary, intuitive notions. \medbreak

{\parindent0pt \textsc{Notion 1: }\textbf{World manifold and world
lines}

We assume that the histories of physical objects can be represented
by lines in an abstract manifold. Properties of this manifold are
to be derived. For simplicity we assume at the moment that the
manifold is of dimension 2 and locally can be mapped to
$\mathrm{R}^{2}$.

\medbreak

\textsc{Notion 2: }\textbf{Assembling at relative rest}

We introduce this intuitive notion. In the world manifold it is
represented by overlapping of two world lines.

\medbreak

\textsc{Assumption 1: }\textbf{Class of proportional clocks and
rulers }

We assume existence of the class of proportional \textsl{clocks}
whose counts rates are constant along their world lines when they
assemble at rest. We also assume existence of the class of
proportional \textsl{rulers} whose world lines of graduation points
are such correlated, that when one pair assembles at rest then also
the other pairs assemble at rest. We assume that the members of the
classes stay proportional each time they assemble at rest.
(\textbf{Homogeneity 1}) If two clocks/rulers are proportional to
the third, they are proportional to each other (class of
equivalence).

\medbreak

\textsc{Notion 3: }\textbf{Time intervals and lengths}

When clocks scale is fixed (by choosing a reference clock),
proportional clocks measure local time interval along their world
lines. Correspondingly, proportional rulers measure (locally) the
lengths of objects staying at rest with the ruler.

\medbreak

\textsc{Definition 1: }\textbf{Blent velocity}

The notions already collected allows us to define a special kind of
relative velocity: defined for clocks passing rulers. If a clock
passes the length $L$ (length of the ruler) in the time interval
$\Delta t$ (time of the clock) we define the blent velocity of the
clock relatively to the ruler as
\be\label{blent}
u=L / \Delta t.
\ee
Choosing orientation of the ruler, we can define positive and
negative velocities.

\medbreak

\textsc{Assumption 2: }\textbf{Relativity 1}

Take two sets, each containing one ruler and one clock assembled at
rest. If the systems pass each other, the values of blent
velocities of these systems are equal.

}
\bigbreak

Our aim will be to construct the algebra of velocities. Here we
will need the last but not least assumption:

\bigbreak

{\parindent0pt

\textsc{Assumption 3: }\textbf{Relativity 2}

Locally the composition of blent velocities depends only on the
relative states of motions.

}

\bigbreak

The last assumption means that having three bodies meeting
somewhere, it is enough to know two relative blent velocities to be
able to calculate the third one.

The Assumption 3 was formulated locally. It means that we have
restricted ourselves to only such cases that fulfill the following
assumption:

\medbreak

{\parindent0pt

\textsc{Assumption 4: }\textbf{Locality}

We consider only pairs of such small rulers for which blent
velocities of separate fragments are equal to each other. It is
equivalent to the postulate of inertial frames in the standard
derivation. Instead of locality we can consider only special states
of relative motions.

}

\medbreak

Equipped with these notions and assumptions, we can sketch the key
picture of our derivation. \medbreak

\begin{picture}(40,55)
\includegraphics{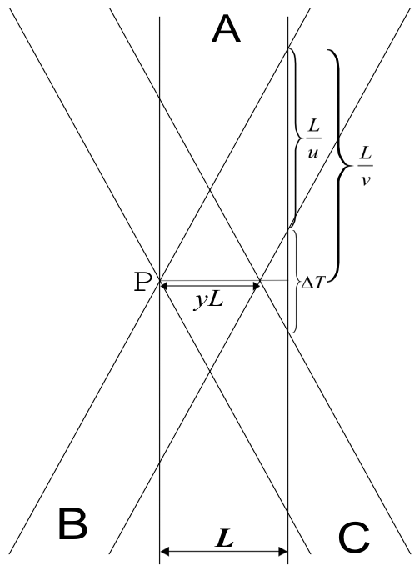} \put(0,-14){\centerline{{\bf Fig. 1}}}
\end{picture}

\section{Key picture}

Take three equal rulers \verb"A", \verb"B" and \verb"C". Let
\verb"B" pass \verb"A" with blent velocity \textsl{u} (to the
right) and let \verb"C" pass \verb"A" with blent velocity
\textsl{-u} (to the left). Let left ends of the rulers meet at some
point \texttt{P}. Then we have three generic cases concerning their
right ends.

{\parindent0pt

\textsc{Case I: }\textbf{(Lorentz case - shown at the picture)}

Right end of \verb"A" meets first the right end of \verb"C" and
then meets the right end of \verb"B"

\medbreak

\textsc{Case II: }\textbf{(Galilean case)}

The three right ends meet at one point.

\medbreak

\textsc{Case III: }\textbf{(Euclidean case)}

The order is opposite to case I.
}

\bigbreak

Thus the cases distinguish by the time interval $\Delta T$
(measured by the clock fixed at the right end of \verb"A") between
right ends meetings of \verb"A"-\verb"C" and \verb"A"-\verb"B".
Choosing a convention we can say that this interval is positive in
the case I and equals zero or is negative in the remaining two
cases respectively.

\medbreak

{\parindent0pt

\textsc{Conclusion 1: }\textbf{Synchronization of clocks}

Observe that the described construction can serve as a procedure of
clock synchronization. In fact, having clocks fixed at the ends of
\verb"A", we can synchronize them in such a way that they show the
same time at \texttt{P} and at half time interval of $\Delta T$.

}

\medbreak

This synchronization works in all three cases! However, we restrict
our considerations here only to the Case I.

\medbreak

Having synchronized clocks, we can define \textbf{standard
velocity} \textsl{v} in a natural way. It is easy to perceive, that
for our picture, eg. for the velocities \textsl{v} and \textsl{u}
of \verb"B" relatively to \verb"A", we have the relation
\be\label{vel}
\frac{1}{v}=\frac{1}{u}+\frac{\Delta T}{2L}
\ee
where $L$ is the length of rulers.

\medbreak

 {\parindent0pt

\textsc{Conclusion 2: }\textbf{Lorentz contraction}

It is also worth to observe, that once the clocks of \verb"A" are
synchronized, we can define the length of moving body: the length
ascribed to it as a distance between its ends at world points
regarded as simultaneous in the clock synchronization of \verb"A".
Thus {\bf in the synchronized reference system of \verb"A"} we can
ascribe the length $L^\prime$ to the ruler of the length $L$ moving
with velocity $v$:
\be\label{contract}
L^\prime = yL
\ee
where $y$ is a contraction factor.}

\medbreak

We will show that our former assumptions are sufficient to derive
the dependence $y(v)$. (Independence of $L$ follows from assumption
4.)

\medbreak

The contracted length of \verb"B" (or \verb"C") is the distance
between point \texttt{P} and the point of \verb"B" and \verb"C"
right ends meeting. So we get
\be\label{ygrek}
yL=\frac{v}{u}
\ee

In order to deriver the dependence $y(v)$ it is convenient to
consider a series of subsequent "additions" of equal velocities.

\section{Sequence of boosts}

Let us consider a question how the velocity and the contraction of
\verb"C" with respect to \verb"B"  depend on $y$ and $v$.

Consider a "boosted" picture.

\begin{picture}(40,75)
\includegraphics{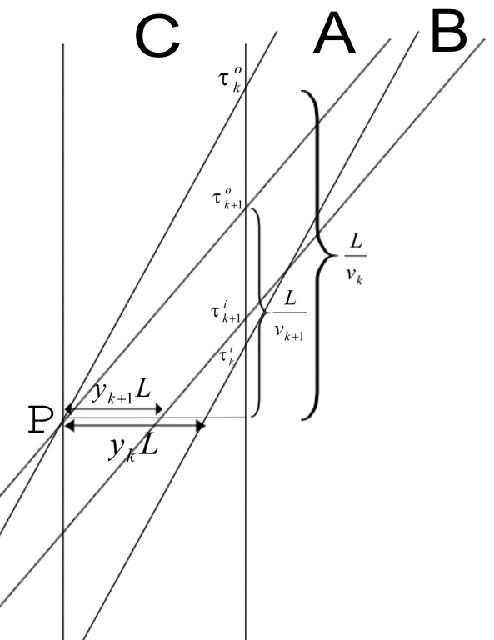} \put(0,-1){\centerline{{\bf Fig. 2}}}
\end{picture}

\bigskip

We have added indexes \textsl{k} for quantities describing the
"old" motion with blent velocity \textsl{u} and indexes
\textsl{k+1} for quantities describing the "new" motion of \verb"C"
relatively to \verb"B" with the "doubled" relative velocity.

\medskip

Going back to figure 1 we can examine that there hold the following
relations between time intervals:
\be\label{intervals1}
v_k(\tau_k^o - \tau_{k}^i) = y_k L
\ee
\be\label{intervals2}
\frac{\tau_k^o - \tau_{k+1}^i}{\tau_k^o - \tau_{k}^i}= \frac{y_k
L}{L}=y_k.
\ee

The symmetry between the "forward" and "backward" motion at figure
1 implies that:
\be\label{intervals3}
\tau_k^o - \tau_{k+1}^o = \tau_{k+1}^o - \tau_{k+1}^i.
\ee
Combining these relations we get
\be\label{intervals4}
\tau_k^o - \tau_{k+1}^o = \frac{1}{2}\frac{L}{v_k}y_k^2.
\ee

If clocks of \verb"C" are synchronized in such a way that
\textsc{C}-time of \texttt{P} is equal 0 then:

\be\label{intervals5}
\tau_k^o=\frac{L}{v_k}, \qquad\qquad \tau_{k+1}^o=\frac{L}{v_{k+1}}
\ee

and we get

\be\label{velo}
\frac{1}{v_{k+1}}=\frac{1}{v_{k}}(1-\frac{1}{2}y_k^2).
\ee

We also get that

\be\label{seria}
y_{k+1}=\frac{y_k^2}{2-y_k^2}.
\ee

Then we can use obtained $v_{k+1}$ and $y_{k+1}$ to construct next
step of iteration and so on.

\section{Results}

The series given by (\ref{seria}) decrease to zero if started from
contraction $y$ smaller than one:
\be\label{zero}
\lim_{k\rightarrow\infty}y_k=0.
\ee

(It can be shown, when we observe that the series (\ref{seria}) is
restricted from above by the series $y_{k+1}=y_k^2$.)

We goes to the following conclusions:

\medskip

{\parindent0pt
\textsc{Conclusion 3: }\textbf{Maximal speed}

By direct insertion one can verify that there holds the relation
\be\label{light}
\frac{v_k^2}{1-y_k^2}=\frac{v_{k+1}^2}{1-y_{k+1}^2}=\mathrm{const}=c^2.
\ee
The new introduced constant $c$ is the maximal relative velocity
which we can obtain combining relative motions (what is clear from
(\ref{light}) and from (\ref{zero})). }

\medskip

The relation (\ref{light}) was obtained for the series
(\ref{seria}). But it is easy to extend it to all relative motions
obtained from "adding" and "subtracting" any motions from the
series and to whichever motion obtained from them.

\medskip

{\parindent0pt \textsc{Conclusion 4: }\textbf{Key result: Lorentz
contraction factor}

Thus we can generally rewrite (\ref{light}) in the form:
\be\label{lorentz}
y=\sqrt{1-\frac{v^2}{c^2}}.
\ee
This is our desired relation between the velocity and the
contraction factor. }

\medskip

Having established relation (\ref{lorentz}) it is straightforward
to derive standard Lorentz group of transformations between moving
systems of rulers and clocks and the Lorentz algebra of velocities
(repeating in fact the Lorentz path). It restores all local
Minkowski space structure.

\section{Conclusions}

We have shown in the simple two dimensional case that the very
natural physical assumptions on the existence of proportional
clocks and proportional rulers, together with homogeneity and
relativity assumptions, allows us to rediscover local Minkowski
structure of space-time. The structure of Minkowski space can be
derived as one of three generic cases (containing also Galilean and
Euclidean case). The signals of light had no special role in our
derivation. In fact the existence of light was even not mentioned.
The parameter of maximal velocity appears as a consequence of
Lorentz contraction. The fact of contraction is treated as a
physical phenomenon distinguishing Minkowski case from the two
other generic cases. No special form of Lorentz contraction was
assumed. It was derived from fundamental assumptions and the
maximal speed appears in the derivation as the only free parameter
whose concrete value has to be fixed from experiment. The above
derivation opens some room for speculations on a nontrivial physics
of the speed of light. In fact, we have not assumed that our
maximal speed is the speed of light. This conclusion - if valid -
has to be derived from physical considerations (eg. from
considerations of the speed of interactions that makes our rulers
and clocks really proportional). But such identification in
principle can be also rejected.

\end{document}